\documentclass[11pt,twoside]{article}

\usepackage{asp2014}

\aspSuppressVolSlug
\resetcounters

\begin{document}

\title{From ESPRESSO to the future -- Analysis of QSO spectra with the Astrocook package}

\author{Guido Cupani, Giorgio Calderone, Stefano Cristiani, Valentina D'Odorico, and Giuliano Taffoni} 
\affil{INAF--Osservatorio Astronomico di Trieste, Trieste, Italy; \email{cupani@oats.inaf.it}}

\paperauthor{Guido Cupani}{cupani@oats.inaf.it}{0000-0002-6830-9093}{INAF}{Osservatorio Astronomico di Trieste}{Trieste}{}{I-34143}{Italy}

\begin{abstract}
The ESPRESSO instrument, to be commissioned in the next months at the ESO VLT, is bound to became a landmark in the field of high-resolution optical spectroscopy, both for its ground-breaking science objectives (search for Earth-like exoplanets; measure of a possible variation of fundamental constants) and for its novel approach to data treatment. For the first time for an ESO instrument, scientific information will be extracted in real time by a dedicated Data Analysis Software (DAS), which includes several interactive workflows to handle the typical analysis cases in stellar and QSO spectroscopy. 
Data analysis tools in the oncoming ELT era will face very demanding requirements from compelling science case, such as the Sandage Test:  the need of handling larger data sizes with a higher degree of accuracy, and the possibility to compare observations and simulated data on the fly. To this purpose, we are currently porting the solutions developed for ESPRESSO to a wider framework, integrating the algorithms within a full-fledged set of Python modules. The project, named ``Astrocook'', is aimed to provide a set of high-level, instrument-agnostic procedures to automatically extract physical information from the data.
\end{abstract}


\section{The era of precision spectroscopy}

Astronomical spectroscopy is rapidly evolving into a precision science. The possibility to acquire visible spectra of distant objects -- such as medium-to-high-redshift QSOs 
at high resolution and with a wavelength accuracy below 1$\,\mathrm{m\,s^{-1}}$ 
is opening exciting opportunities in 
the field fundamental physics, e.g.~the possibility to determine a variation of the fundamental constants \citep{2017RPPh...80l6902M}, or to measure the accelerated expansion of the universe from a redshift drift of distant sources \citep{1962ApJ...136..319S,2008MNRAS.386.1192L}.
Several instruments are being conceived and realized to meet unprecedented requirements in terms of stability and repeatability of the observations; 
one is ESPRESSO \citep{2013Msngr.153....6P}, a ultra-stable spectrograph for the ESO Very Large Telescope (VLT), which was intended since its inception as a precursor of the future high-resolution spectrograph \citep{2014SPIE.9147E..23Z} for the ESO Extremely Large Telescope (ELT). ESPRESSO is the first ESO instrument to be equipped with a dedicated Data Analysis Software or DAS \citep{2012SPIE.8448E..1OD,2015ASPC..495..289C,2016SPIE.9913E..3RD}, which is included in the instrument package together with the Data Reduction Software or DRS. This article describes the lessons learned in developing the ESPRESSO DAS, and presents first implementation of the new ``Astrocook'' Python package, which is meant to pave the way towards the next generation of data processing systems \citep{2016SPIE.9910E..2FC}.  

\articlefigure{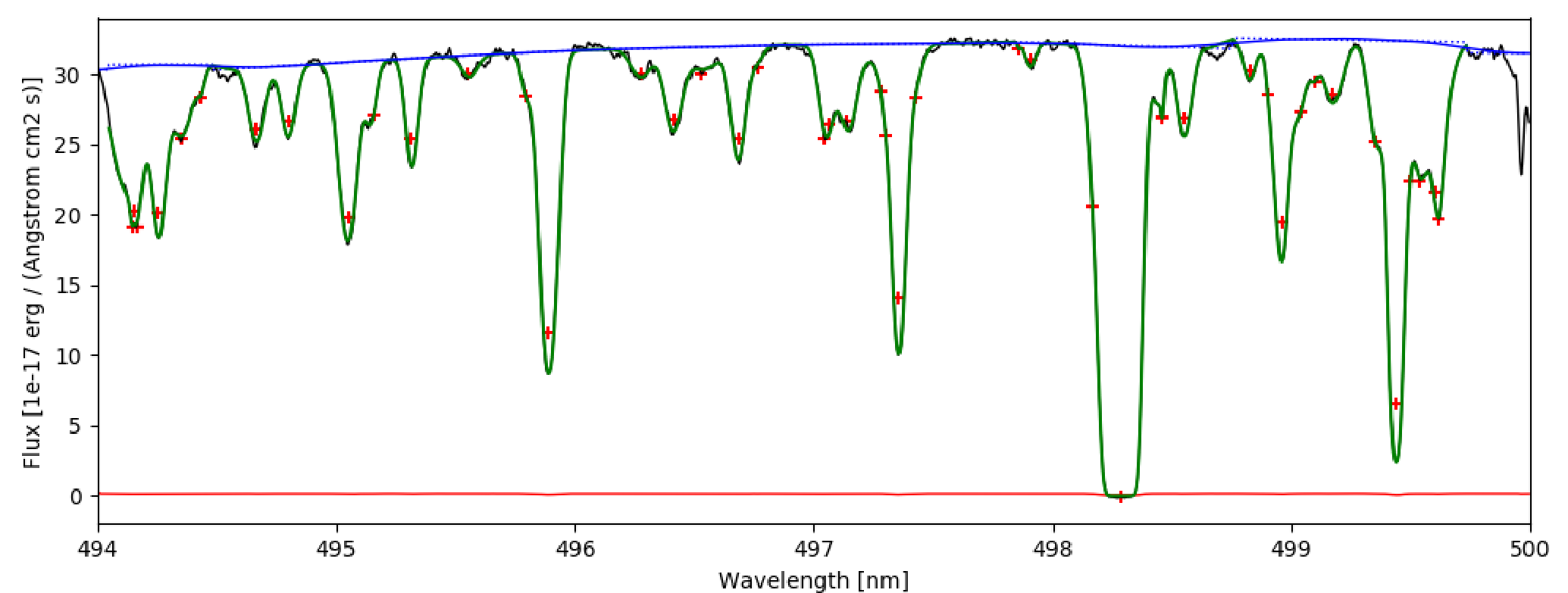}{cont}{A portion of the Lyman-$\alpha$ forest of QSO J0515-4410 (black line: flux density; red line: error on flux density) as fitted by the Astrocook package. Information obtained from automated Voigt-profile fitting (green line) of the detected lines (red crosses) is used to locally adjust the guess continuum (dotted blue line); the final continuum is determined by smoothing the result after iteration (blue line).}

\section{The ESPRESSO Data Analysis Software}

ESPRESSO has been purposely designed as an end-to-end ``science machine'', which is fed photons from the telescope and outputs not just calibrated spectra, but actual astrophysical information about the observed targets \citep{2013Msngr.153....6P}. In the case of QSO spectra, such information includes (i) the determination of the emission continuum of the QSO; (ii) the detection of individual absorption features (lines) that can be modeled as the superposition of different Voigt-profile components; and (iii) the interpretation of such lines in terms of absorption systems, including different neutral hydrogen and metal transition at the same redshift, with a given column density and thermal broadening. Emission and absorption features are entangled in a way that requires a simultaneous treatment; most significantly, neutral hydrogen lines giving rise to the so-called Lyman forest require to be fitted with respect to a reference continuum emission, but in turn they provide information about the opacity $\tau_\mathrm{HI}$ as a function of redshift, which can be used to refine the estimation of the continuum itself. Only an iterative approach can cope with such situation and obtain a proper fit of both continuum and lines \citep[see also Sect.~\ref{astrocook}]{2016SPIE.9913E..3RD}. 

In developing the ESPRESSO DAS, much effort was put in having the software modules mimic what human observers would do ``by hand'' or ``by eye''. The human brain is extremely powerful at detecting spectral features and mentally subtract them to let the underlying pattern emerge (e.g.~removing a line system to guess the shape of the original emission), but it lacks the capability to handle large quantities of data and is generally prone to subjective bias. The DAS procedures reproduce some manual operations (e.g.~using a detected C\textsc{iv} doublet as a starting point to identify different metal transitions at the same redshift, or adding components to a line system where the fit residuals are high, to improve the model) and perform then automatically along the whole spectrum, or sequentially across a catalogue of spectra. 
The DAS code is written mostly in ANSI C, taking advantage of the Common Pipeline Library \citep[CPL,][]{2004SPIE.5493..444M} implemented for the whole ESO Data Flow System. Code modules (``recipes'') can be launched as stand-alone plugins, or invoked through the ESO Reflex interface \citep{2013A&A...559A..96F}, which can be used to implement different pipelines for cascade execution (``workflow'') through a graphical user interface. A fixed workflow has been set up for the analysis of QSO spectra, allowing the users to inspect the recipe products and interactively set up the recipe parameters through a set of Python scripts. The code is distributed under the GNU General Public License; its first public release will be issued after the commissioning of ESPRESSO (early 2018).

The instrument-specific strategy adopted by the DAS has of course some limitations: (i) being tailored to the output of the DRS, it is not easily applied to spectra from other instruments; (ii) its workflow execution is limited by the capability of the ESO Reflex environment, which may not be suited to all situations; (iii) it doesn't allow for finer configuration below the recipe level: recipes are provided as black boxes not meant to be modified by the users; only recipes parameter can be tuned at will. While such features do not impact in the DAS ability to properly handle ESPRESSO spectra, they may hinder the prospective generalization of the DAS model to the future generation of high-resolution spectrographs. To overcome the issue, the Astrocook package was developed.

\section{The Astrocook package}\label{astrocook}

Astrocook is a new Python package to analyze quasar spectra. The name was originated by the tagline ``a thousand recipes to cook a spectrum'', which is meant to emphasize the versatility of the tool. The project is still in its infancy, but a working copy of the package can be downloaded for testing from its GitHub webpage (\url{https://github.com/DAS-OATs/astrocook}). The idea is to create a general representation of a spectrum (and its associated metadata) in the form of a Python object, and to develop an instrument-agnostic set of procedures that the users may invoke through simple scripts, designing their own workflows with the desired level of control in the sequence of operations and in the iteration schemes. 

The algorithms developed for the ESPRESSO DAS were divided into atomic constituents, thus breaking the recipe-as-a-black-box constraint of the original code; in addition, several new solutions were added, to improve functionality and address specific analysis requirement. Python 3 was chosen as a development language due to its current prevalence in the astrophysical data analysis community; we took advantage, in particular, of the NumPy \citep{van2011numpy} and Astropy \citep{2013A&A...558A..33A} packages for general data handling, and of the LMFIT package \citep{newville_2014_11813} for non-linear least-square minimization of Voigt profiles.

As an example of the difference between the DAS and the Astrocook we discuss the algorithm to fit the emission continuum by removing the absorption lines. In the DAS, this task is performed by the recipe \texttt{espda\_fit\_qsocont}, which proceeds as follows: (i) the spectrum is split into a blue and a red part, using the Lyman-$\alpha$ emission as a demarcation; (ii) in the red part, previously-detected absorption lines are fitted all at once with respect to an interpolated continuum; (iii) in the blue part, lines are fitted iteratively in spectral chunks, from the strongest to the weakest, with respect to a guess continuum which is determined by taking into account the residual $\tau_\mathrm{HI}$ of the lines yet to be fitted; (iv) the final continuum is determined with a cubic spline interpolation, after the fitted lines are removed from the spectrum. Such procedure is effective in simultaneously interpreting the emission and absorption features, but does not accommodate for further iterations and is subject to local failure which require an ad hoc treatment. Conversely, Astrocook provides a set of methods to perform the operations separately: (i) estimate the guess continuum by detecting and masking the absorption lines and applying a suitable smoothing technique to the masked spectrum (e.g.~a Savitzky-Golay filter); (ii) identify the absorption features as Lyman-$\alpha$ lines or metal doublets and organize them into groups to be fitted together; (iii) fit the absorption features with respect to the guess continuum, automatically adding Voigt component to improve the goodness of fit; (iv) refine the guess continuum based on the information extracted from the line, possibly including the continuum normalization as a free parameter in the line fit; etc. An example of the results is shown in Fig.~\ref{cont}. Each module may run independently or taking prior information from the products of other modules, providing an extreme liberty in designing the procedure. This is precisely the kind of flexibility which is required to tune the code to specific analysis cases while maintaining control over the reproducibility and repeatability of the execution. 

The scientific exploitation of Astrocook is ongoing. The package is meant to collect contributions from the QSO data analysis community as large; a porting of the QSFit package for emission line fitting \citep{2017MNRAS.472.4051C} is currently being implemented, and several additions (ranging from the template-based flux calibration to the creation of mock spectra for evaluating the statistical significance of the results) are under consideration. The experience of the DAS ``on the field'', when ESPRESSO comes into operation, will motivate the further development of the package in the years leading to the ELT first light.





\bibliography{O4-5}

\end{document}